\documentclass[12pt,a4paper]{article}

\usepackage{amsmath}
\usepackage{amsfonts}
\usepackage{amssymb}
\usepackage{graphicx}
\usepackage{color}

\newcommand{\be}{\begin{eqnarray}}
\newcommand{\ee}{\end{eqnarray}}

\newcommand{\A}{\mathrm{A}}

\newcommand{\C}{\mathrm{C}}

\newcommand{\T}{\mathrm{T}}

\newcommand{\w}{\mathrm{w}}

\renewcommand{\i}{\mathrm{i}}
\renewcommand{\j}{\mathrm{j}}
\renewcommand{\k}{\mathrm{k}}
\renewcommand{\v}{\mathrm{v}}
\renewcommand{\u}{\mathrm{u}}

\newcommand{\p}{\mathrm{p}}

\newcommand{\Tr}{\mathrm{Tr}}

\newcommand{\Ch}{\mathrm{Ch}}

\newcommand{\sign}{\mathrm{sign}}
\newcommand{\AII}{\mathrm{AII}}

\newcommand{\BH}{\mathbf{H}}

\newcommand{\Bp}{\mathbf{p}}
\newcommand{\BR}{\mathbf{R}}

\newcommand{\CC}{\mathbb{C}}

\newcommand{\ZZ}{\mathbb{Z}}

\renewcommand{\bar}{\overline}
\renewcommand{\hat}{\widehat}
\renewcommand{\tilde}{\widetilde}

\newcommand{\half}{\frac{1}{2}}

\renewcommand{\(}{\left(}
\renewcommand{\)}{\right)}

\begin{document}
\begin{center}
{\Large \textbf{Simple Models for All Topological Phases}}\\
\bigskip
\smallskip
\centerline{Tong Chern$^\dagger$}
\smallskip
\centerline{\it School of Science, East China Institute of Technology, Nanchang 330013, China}
\medskip \centerline{\it $^\dagger$ 1643399509@qq.com}
\end{center}
\begin{abstract}
We construct simple models for all topological phases of free fermions.
These explicit models can realize all the nontrivial topological phases
(with any possible topological invariant)
of the periodic table.
Many well known models for topological insulators
and superconductors are special cases of our general constructions.

\end{abstract}
\tableofcontents

\bigskip
\bigskip

\section{Introduction}

~~~~~There has been a surge of interest
in the study of topological phases of free fermions\cite{TI1}\cite{TI2}\cite{TI3}.
These free fermion systems can be classified according to the topology of ground state wave functions
\cite{kane1}\cite{kane2}\cite{kane3}\cite{3d1}\cite{3d2}.
In this classification, one groups single particle
Hamiltonians that can be smoothly deformed into each others without closing
the bulk gap as a homotopical class.
The systematical classifications show a regular pattern
and can be arranged into a periodic table\cite{pt1}\cite{pt21}\cite{pt2}\cite{pt3}.
In this periodic table, the topological phases of free fermions are characterized by three numbers,
the symmetry class in the Altland-Zirnbauer(AZ) classification\cite{az1}\cite{az2},
the dimensionality, and the $\ZZ$ or $\ZZ_2$ valued topological invariant.
These three numbers are not independent, but are related by the Bott periodicity.

Many models for the two or three dimensional topological phases (with lowest nonzero topological invariants)
of the periodic table have been known
(see \cite{TI1}\cite{TI2}\cite{TI3} and references there in).
Some of these models have been realized in real materials,
such as the BHZ model for HgTe/CdTe quantum wells (proposed by \cite{BHZ}, and realized by \cite{Konig}),
and the low energy effective model for $\mathrm{Bi_2Se_3}, \mathrm{Bi_2Te_3}$ and $\mathrm{Sb_2Te_3}$
 (proposed by\cite{TI}, and realized by \cite{Chen}\cite{Hsieh}), etc.
But a systematical model construction for all nontrivial
topological phases seems to be lacking (some recent papers that focus on
 or partially focus on model building include \cite{duan}\cite{chern}\cite{ryu2}\cite{Fu}\cite{ryu}).

Thus, one of the remaining problems is, given an arbitrary spatial dimensions,
given an arbitrary symmetry class, and especially given an arbitrary topological invariant,
whether or not one can construct an explicit model that can realize
the corresponding topological phase (with the numbers of bands as few as possible)?
The main purpose of the present paper is to solve this problem, through simple, but
explicit and concrete model constructions.

In the present paper, we reduce this problem to the problem of building nontrivial
models for all the nontrivial topological phases of chiral classes
(half of the ten AZ symmetry classes are chiral)
and with integer topological invariants at the same time.
This can be achieved by utilizing two simple ideas.
The first idea is to construct the general model
with topological invariant $n$ as the \emph{powers of $n$ model} of a basic model.
This basic model has lowest nonzero topological invariant.
The second simple idea is to use the Clifford algebras over the field of real numbers.
And we will use both the abstract structure of Clifford algebras and their irreducible representations (IR)
(both the complex IR and the real IR).

As it is well known, Clifford algebras and their representations
have been used to the classifications of topological insulators and superconductors
\cite{pt1}\cite{pt21}\cite{pt3},
and to the model constructions of quantum spin Hall insulators \cite{zhang}\cite{kane1}.
Moreover, Clifford algebras are also important in the recent many works concerning Dirac semi-metal
\cite{Nagaosa}\cite{Kanedirac1}\cite{Kanedirac2}\cite{witten}.
But there are two mainly differences between our utilizations
to Clifford algebras and the previous. The first difference is that we will mainly use the Clifford algebras
with the generators square to minus one rather than square to plus one.
The second difference is that, in the previous utilizations,
people either mainly use the abstract algebraic structure of Clifford algebras\cite{pt21},
or mainly use their gamma matrix representations
\cite{zhang}\cite{kane1}\cite{pt1}\cite{pt3}\cite{witten},
while in our utilizations to Clifford algebras,
both the abstract algebra and their irreducible representations are important.

After building models for all the topological phases of chiral classes and with integer topological invariants.
We can further build models for non-chiral classes that have integer topological invariants, by
lifting a $d$ dimensional chiral model to a corresponding $d+1$ dimensional model with the same topological invariants.
In this dimensional lifting, the chiral symmetry is broken, thus we will get a $d+1$ dimensional non-chiral model.
This procedure is originally proposed in \cite{pt3}. A notable change
in this dimensional shifting is that the Clifford algebras with the generators square to minus one
are shifted to the Clifford algebras with generators square to plus one. Thus,
the Clifford algebras associated with the model constructions for chiral classes and for non-chiral
classes are two different kinds of Clifford algebras.

As for the model constructions for $\ZZ_2$ topological insulators and superconductors,
we will use the well known dimensional reduction procedure\cite{pt1}\cite{pt3}.
By using this procedure, one can get two $\ZZ_2$ descendants of $d-1$ and $d-2$ dimensions respectively,
from a parent $d$ dimensional model with integer topological invariants.
These descendants are in the same symmetry classes with their parent model.
We will illustrate this procedure by \emph{theoretically deriving} the BHZ model and the three dimensional model
for $\mathrm{Bi_2Se_3}, \mathrm{Bi_2Te_3}$ and $\mathrm{Sb_2Te_3}$,
from a simple four dimensional parent model that constructed by the dimensional lifting procedure.

The present paper will mainly deal with the eight real symmetry classes of the periodic table,
since this is the most complicate and interesting situation.
But we will also discuss the model constructions for the two complex classes.
And we found that, different with the real cases,
in the complex situation, the most natural Clifford algebras that we should use
are the Clifford algebras over the field of complex numbers
rather than over the field of real numbers. Finally, we will demonstrate that,
both for the real cases and for the complex cases,
the periodicity of the periodic table can be naturally
interpreted as the Bott periodicity of the corresponding Clifford algebras
\cite{atiyah}.

This paper is organized as follows. In section 2, we will focus on the model constructions
for three dimensional topological insulators and superconductors that have integer topological invariants.
Our main purpose is to illustrate the basic ideas of the present paper, by using this relatively simple situation.
In section 3, we will summarize some relevant mathematical facts concerning the Clifford algebras.
Also in this section, two simple mathematical lemmas are developed.
These lemmas are important for the model constructions in general situations.
The main part of this paper is section 4. In this section,
we will explicitly construct models for all the topological phases
that have both chiral symmetry and integer topological invariants.
We will give our detail constructions in three types,
discussed in section 4.2, section 4.3 and section 4.4 respectively.
Section 4.1 is devoted to some general aspects of the model constructions for the chiral classes.
In section 4.4, we will relate the Bott periodicity of
the eight real classes to the Bott periodicity of the Clifford algebras.
In section 5, we extend our model constructions to all the topological phases
(with or without chiral symmetry) that have integer topological invariants.
Section 6 is devoted to the model constructions for all $\ZZ_2$ topological phases.
In this section, as illustrative examples, we will give a \emph{theoretical derivation} to the well known three dimensional model
for $\mathrm{Bi_2Se_3}, \mathrm{Bi_2Te_3}$ and $\mathrm{Sb_2Te_3}$, and to the BHZ model for quantum spin Hall insulators.
Finally, in section 7, we extend our model constructions to the remaining two complex symmetry classes (class A and class AIII).
This complete our model constructions for all the topological phases.

 \section{Models for $\ZZ$ Topological Insulators and Superconductors in Three Dimensions}

 ~~~~~In this section, we will construct models for three dimensional $\ZZ$ topological insulators and superconductors
 (i.e. the class DIII and class CI, since we are focusing on the real cases of the periodic table).
 We want to use this relatively simple situation to illustrate the basic ideas behind the present paper.
In this situation, the general theory of
 Clifford algebras and their representations are not necessary.
This greatly reduce the degree of mathematical abstraction,
so that the basic ideas behind our constructions can be seen more clearly.

The first simple idea behind our constructions, as has been mentioned in the introduction,
is to build general models as \emph{the powers of a basic model}.
This basic model carries minimum nonzero topological invariant.

The second simple idea is to use the Clifford algebras with all the generators square to $-1$.
The importance of Clifford algebra is that, we can use it to build a large class of models, these models have the
particular property that arbitrary powers of them will be in the same symmetry class.
Thus, it is exactly the Clifford algebras that make us can fulfill our first simple idea.

In the following two subsections, we will make these two ideas mathematically accurate,
by using the specific model constructions for three dimensional $\ZZ$ topological insulators (and superconductors).

\subsection{Three Dimensional Model Construction in General}

~~~~~As one can see (from the periodic table) that, at odd spatial dimensions, such as three dimensions, the $\ZZ$ topological insulators and superconductors
have chiral symmetry $S$, with $S^2=1$. $S$ is anticommuting with the Bloch Hamiltonians (or the BdG Hamiltonians).
Hence one always can choose a particular basis by diagonalizing the chiral symmetry $S$,
in this basis the Bloch Hamiltonian(or BdG Hamiltonian) is off diagonal£¬
\be \left(\begin{array}{ccccc} 0 & \u(\Bp)
\\ \u^{\dagger}(\Bp) & 0
\end{array}\right).\ee
We will denote this Hamiltonian as the model $\{\u(\Bp)\}$. It is easy to see that the energy eigenvalue $E(\Bp)$
of $\{\u(\Bp)\}$ must satisfy an equation like $\u^{\dagger}(\Bp)\u(\Bp)\phi=E^2\phi$, here $\phi$ is a column vector.
Hence if our model is fully gapped (as required in topological insulators),
then $E(\Bp)\neq 0$. Then the determinant of matrix $\u^{\dagger}\u$ and thus the determinant of matrix $\u(\Bp)$
are always nonzero at the whole momentum space.
Consequently, in this case the matrix $\u(\Bp)$ is always invertible.

Now we can define the \emph{powers of $n$($n>0$) model} of model $\{\u(\Bp)\}$ as follows,
\be \left(\begin{array}{ccccc} 0 & \u^n(\Bp)
\\ (\u^{\dagger})^n & 0
\end{array}\right).\ee
We will denote this powers of $n$ model as $\{\w(\Bp)\}$, with $\w(\Bp)=\u^n(\Bp)$.
We can proof that the topological invariant of model $\{\w(\Bp)\}$ is just $n$ times the topological invariant of model $\{\u(\Bp)\}$.

To give a proof, we firstly need to give the definition of the topological invariant
of an arbitrary three dimensional chiral model. In fact this is well known \cite{pt2}.
For example, for the three dimensional model $\{\u(\Bp)\}$ ($\u(\Bp)$ is invertible),
the corresponding topological invariant $\nu_3(\u)$ is the winding number
\be\nu_3(\u)=\int_{\Bp}\Ch_3(\u),\label{ch3}\ee
here the integration is over the whole momentum space, with
\be\Ch_3(\u)=\frac{1}{24\pi^2}\Tr(\u^{-1}d\u\wedge\u^{-1}d\u\wedge\u^{-1}d\u).\label{ch31}\ee
Here, we use the symbol $\Ch_3$ to denote the three form of the winding number,
the reason behind this convention is explained in \cite{chernK}.
And we have slightly generalized the original definition \cite{pt2} for the topological invariant $\nu_3$.
Our definition is equivalent to the original one,
but the present definition is more straightforward,
since we do not require the matrix $\u(\Bp)$ to be a unitary matrix.
In fact, the matrix $\u(\Bp)$ in our formula (\ref{ch31})
is just the off diagonal submatrix of model $\{\u(\Bp)\}$.

Now given two models $\{\u_1(\Bp)\}$,$\{\u_2(\Bp)\}$,
one can easily proof that the difference between the differential form
$\Ch_3(\u_1\u_2)$ and the differential form $\Ch_3(\u_1)+\Ch_3(\u_2)$ is an exact form.
Thus, in the sense of de Rham cohomology, we have
\be\Ch_3(\u_1\u_2)=\Ch_3(\u_1)+\Ch_3(\u_2).\ee
Consequently, $\nu_3(\u_1\u_2)=\nu_3(\u_1)+\nu_3(\u_2)$. We then have \be
\nu_3(\w)=\nu_3(\u^n)=n\nu_3(\u).\ee
Thus, to construct a model with arbitrary topological invariants $n$,
we can firstly construct a basic model with $\nu_3(\u)=\pm 1$,
then the powers of $n$ models of this basic model
will automatically have topological invariants $\pm n$.

But, to fulfill this procedure, we still need (this is the key point) to proof that,
the powers of $n$ model $\{\u^n(\Bp)\}$ are in the same symmetry class with the basic model $\{\u(\Bp)\}$.
In generally, if $\u(\Bp)$ is an arbitrary matrix, this can not be true.
But, as we will show, by utilizing the Clifford algebras, we can construct a large class of models,
in these models $\{\u^n(\Bp)\}$ and $\{\u(\Bp)\}$ are indeed in the same symmetry class.

The Clifford algebra that we will use is the Clifford algebra $C_3$ over the field of real numbers.
$C_3$ have only three Clifford generators $e_1,e_2,e_3$, these generators satisfy $e^2_i=-1, e_ie_j+e_je_i=0(i\neq j)$.
We can easily see that, $e_1e_2e_3$ is commuting with all elements of $C_3$,
and $(e_1e_2e_3)^2=1$, hence $e_1e_2e_3=\pm 1$. Thus $C_3$ can be decomposed as a direct sum of two algebras
(with different signs of the $\pm 1$). Both algebras are isomorphic to the algebra of quaternions, $\BH$.
Taking the algebra of $e_1e_2e_3=-1$ as an example, if one set $e_1=\i, e_2=\j, e_3=\k$,
then one will get the algebra of quaternions,
\begin{equation}\i^2=\j^2=\k^2=-1,\ \i\j=-\j\i=\k,\
\j\k=-\k\j=\i,\ \k\i=-\i\k=\j.\label{quaternion}\end{equation}
Here, $\i, \j, \k$ are the three generators of $\BH$.

Thus, in three spatial dimensions, instead of the full mechanism of the Clifford algebras,
we can use the more familiar quaternion algebra to fulfill our model building.
Notably, before our present work, the quaternion algebra has been used to construct models of three dimensional $\ZZ$ topological insulators
(and superconductors)\cite{duan}. Thus there may be some similarities between our constructions and the constructions of \cite{duan}.
But there are also some fundamental differences. Firstly, in the constructions of \cite{duan},
only the algebraic structure of quaternion algebra has been used, while in our constructions
both the abstract algebraic structure and its irreducible representations (including the complex two dimensional
irreducible representation and the four dimensional real irreducible representation) are important.
As one will see, this combination of the abstract algebra $\BH$ and its representation will greatly simplify our constructions.
More importantly, by using the more general Clifford algebras,
our methods can be naturally generalized to arbitrary spatial dimensions.
Moreover, what \cite{duan} considered are the tight binding models, while our present paper will focus on the continuous models
(even if it is quite easy to rewrite our models to the form of tight binding). This is because that the mathematical manipulations
for the continuous models are simpler, and because in the continuous limit, some subtle mathematical mechanisms can be seen more clearly.

As it is well known, the generators $\i,\j,\k$ of $\BH$ can be represented as anti-Hermitian matrixes.
In the situation of complex two dimensional irreducible representation(IR), a convenient choice for
the corresponding representation matrixes is to represent $\i, \k$ as pure imaginary matrixes,
while represent $\j$ as the following $2\times 2$ antisymmetrical real matrix
\be\j=\left(\begin{array}{ccccc}0 & 1
\\ -1 & 0
\end{array}\right).\ee
More specifically, one can choose $\i=i\sigma_x, \j=i\sigma_y, \k=-i\sigma_z$, where $\sigma_x, \sigma_y, \sigma_z$ are the three Pauli matrixes.

To get the four dimensional real IR of $\BH$, one only need to
replace the unit number $1$ as a $2\times 2$ unit matrix, and replace all the imaginary unit $i$
as the following $2\times 2$ antisymmetrical real matrix,
\be \left(\begin{array}{ccccc}0 & 1
\\ -1 & 0
\end{array}\right).\ee
Obviously, all the matrixes of $\i,\j,\k$ in four dimensional real IRs are antisymmetrical real matrixes.

Notably, in our present paper, we will use the same symbols, such as $\i,\j,\k$, to denote both the
abstract algebraic elements and their irreducible representation matrixes. We hope that
this more flexible utilizations to algebraic symbols will not lead to misunderstandings.

Given a specific quaternion $\A=a_0+a_1\i+a_2\j+a_3\k$, $a_0,a_1,a_2,a_3$ are real numbers,
we can define the quaternion conjugation $\bar{\A}$ of $\A$ as
$\bar{\A}=a_0-a_1\i-a_2\j-a_3\k$. When $\A$ is viewed as the representation matrix,
this quaternion conjugation is then identical with the
Hermitian conjugation $\bar{\A}=\A^{\dagger}$.
The modulus of $\A$ is given by $|\A|=\sqrt{\bar{\A}\A}=\sqrt{a^2_0+a^2_1+a^2_2+a^2_3}$.

\subsection{The Models for Symmetry Class DIII and Class CI in Three Dimensions}

~~~~~Now, let us try to construct models for three dimensional Class DIII topological superconductors.
We will begin from the basic model $\{\u(\Bp)\}$, its BdG Hamiltonian $H$ can be written as the off diagonal form
\be H=\left(\begin{array}{ccccc} 0 & \u(\Bp)
\\ \bar{\u}(\Bp) & 0
\end{array}\right).\ee
Here, we take $\u(\Bp)$ as a quaternion $\u(\Bp)=u_0(\Bp)+u_1(\Bp)\i+u_2(\Bp)\j+u_3(\Bp)\k$.
In specifically, we take
\be \u(\Bp)=t(p^2_x+p^2_y+p^2_z)-\mu+\Delta_0(p_x\i+p_y\j+p_z\k)=t\Bp^2-\mu+\Delta_0\p,\label{construction1}\ee
where $t,\mu,\Delta_0$ are real constants.
One can easily get the eigenvalues of $H$
\be E_{\pm}(\Bp)=\pm\sqrt{(t\Bp^2-\mu)^2+\Delta^2_0\Bp^2}.\ee
Obviously, when $\mu\neq 0$, $H$ is fully gapped. Thus one can use formula (\ref{ch3}) to calculate the topological invariant.
When the quaternions are in two dimensional complex IR, the result is
\be\nu_3(\u) &=&\int_{\Bp} \Ch_3(\u)\\
&=& \frac{1}{12\pi^2}\int_{\Bp}\frac{\epsilon_{ijkl}u^{i}du^{j}\wedge du^{k}\wedge du^{l}}{|\u|^4}\\
&=&\deg(\u).\ee
Where $\deg(\u)$ denotes the mapping degree of map $\Bp\rightarrow \u(\Bp)/|\u|$.
Obviously, $\u(\Bp)/|\u|$ is a quaternion with unit modulus,
hence $\Bp\rightarrow \u(\Bp)/|\u|$ is just the map from three dimensional momentum space
to the unit sphere $S^3$ of four dimensional Euclidean space.
If $\mu t<0$, one will get $\deg(\u)=0$, thus our model $\{\u(\Bp)\}$ will be in topologically trivial phase.
While when $\mu t>0$, the well known mechanism of band inversion tells us
$\deg(\u)=-1\cdot \sign(\mu)$.

With the same quaternion $\u(\Bp)$, but if the quaternions are in four dimensional real IR,
the topological invariants will then be doubled, that is to say, in this situation, we will get $\nu_3(\u)=2\deg(\u)$.
The reason is that, in the four dimensional representation, while the algebra is the same,
the rank of the representation matrixes is doubled,
thus we will get an additional factor of two in carrying out the trace in (\ref{ch31}).
But, as we will see, if the quaternions of the same construction $\u(\Bp)$ are in four dimensional real IR rather than in 2d complex IR,
our model $\{\u(\Bp)\}$ will describe the class CI topological superconductors
rather than the class DIII topological superconductors.

The powers of $n$ model of our basic model $\{\u(\Bp)\}$ can be denoted as
$\{\w(\Bp)=\u^n(\Bp)\}$, with
\be\left(\begin{array}{ccccc}0 & \u(\Bp)^n
\\ \bar{\u}(\Bp)^n & 0
\end{array}\right).\label{n1}\ee And as we have mentioned, the key point is to demonstrate that
all the models $\{\w(\Bp)=\u^n(\Bp)\}$ (with arbitrary $n$) are in the same symmetry class.
In what follows we will show that, with the specific construction (\ref{construction1}) for the basic model,
and in the complex IR of the quaternions, we can have
the particle hole symmetry $\C$ and time reversal symmetry $\T$ with
$\C^2=1,\T^2=-1$, for all the models of $\{\w(\Bp)=\u^n(\Bp)\}$.
Thus, in this situation, we achieved the model building for the class DIII topological superconductors.

In fact, as one can easily see that, for our basic model $\{\u(\Bp)\}$ given in (\ref{construction1}),
there is $\bar{\u}(\Bp)=\u(-\Bp)$. And since,
in the complex IR, $\j\u=\u^*\j$ (here $*$ stands for complex conjugation),
thus we can see that $\{\w(\Bp)=\u^n(\Bp)\}$ always have particle hole symmetry $\C$ and time reversal symmetry $\T$
(for arbitrary $n$), with
\be \C&=&\left(\begin{array}{ccccc}0 & \j
\\ -\j & 0
\end{array}\right)K\\
\T&=&\left(\begin{array}{ccccc}0 & \j
\\ \j & 0
\end{array}\right)K.\ee
Where $K$ stands for the complex conjugation.
Obviously, $\C^2=1,\T^2=-1$, thus, indeed, our models $\{\u^n(\Bp)\}$
are describing class DIII topological superconductors.
If $\mu t>0$, these models will be in the topologically
nontrivial phases, with topological invariants $n$.

Now, we try to rewrite our models $\{\u^n(\Bp)\}$ in more usual form.
Firstly, for all positive $n$, $\w(\Bp)=\u^n$ is, of course, still a quaternion, hence can be written as
\be \w(\Bp)=E(\Bp)+\Delta(\Bp),\ee
where $E(\Bp)$ is a real number, and $\Delta=\Delta_x\i+\Delta_y\j+\Delta_z\k$ is a pure quaternion.
Obviously, $\bar{\w}(\Bp)=E(\Bp)-\Delta(\Bp)$.
On the other hand, since our basic model $\{\u(\Bp)\}$ satisfy $\bar{\u}(\Bp)=\u(-\Bp)$,
thus $\bar{\w}(\Bp)=\bar{\u}^n(\Bp)=\u^n(-\Bp)=\w(-\Bp)=E(-\Bp)+\Delta(-\Bp)$.
Consequently, $E(\Bp)$ must be an even function of $\Bp$, while $\Delta(\Bp)$ must be a odd function of $\Bp$,
\be E(-\Bp)=E(\Bp),\Delta(-\Bp)=-\Delta(\Bp).\ee

Now to rewrite our constructions $\{\u^n(\Bp)\}$ in a more usual basis,
one can make a unitary transformation
$\left(\begin{array}{ccccc} 1 & 1
\\ 1 & -1
\end{array}\right)/\sqrt{2}$ to the off diagonal Hamiltonian (\ref{n1}), then one has
\be H_{BdG}=\left(\begin{array}{ccccc} E(\Bp) & \bar{\Delta}(\Bp)
\\ \Delta(\Bp) & -E(\Bp)
\end{array}\right).\label{bdg1}\ee
In the present basis, the particle hole symmetry $\C$ and time reversal symmetry $\T$ act as
\be \C&=&\left(\begin{array}{ccccc}0 & \j
\\ -\j & 0
\end{array}\right)K\\
\T&=&\left(\begin{array}{ccccc}\j & 0
\\ 0 & -\j
\end{array}\right)K.\ee
And the second quantized mean field Hamiltonian $\hat{H}$ can be written as
\be\hat{H}=\half \int_{\Bp}\Psi^{\dagger}(\Bp)H_{BdG}\Psi(\Bp),\ee
where the Nambu spinor $\Psi(\Bp)$ is given by,
\be\Psi(\Bp)=(c_\uparrow(\Bp), c_\downarrow(\Bp), c^{\dagger}_\downarrow(-\Bp), c^{\dagger}_\uparrow(-\Bp))^{T}.\ee

If we take $\i=i\sigma_x, \j=i\sigma_y, \k=-i\sigma_z$, then we can rewrite the BdG Hamiltonian (\ref{bdg1}) more explicitly.
In the case of $n=1$, it is (here we have make a simple unitary transformation to get rid of some inessential imaginary unit $i$)
\be H_{BdG}=\left(\begin{array}{ccccc} E(\Bp) & 0 & -\Delta_0p_z & \Delta_0p_-\\
0 & E(\Bp) & \Delta_0p_+ & \Delta_0p_z\\
-\Delta_0p_z & \Delta_0p_- & -E(\Bp)& 0\\
\Delta_0p_+ & \Delta_0p_z & 0 & -E(\Bp)
\end{array}\right),\label{tsc}\ee
here $p_{\pm}=p_x\pm ip_y$, $E(\Bp)=t\Bp^2-\mu$.
As one can see that, (\ref{tsc}) is identical with the well known model
for triplet pairing three dimensional topological superconductor \cite{TI2}.
When $\mu>0$(of course in the case of $t>0$),
this model (\ref{tsc}) will transit to a topologically nontrivial phase,
with the topological invariant $1$, these results are of course well known.

Given the same construction with the basic model given by (\ref{construction1}),
but if we now take the four dimensional real IR for the quaternions.
Then the particle hole symmetry $\C$ and time reversal symmetry $\T$ of (\ref{n1}) will be
(noticing that $\i,\j,\k$ are now realized as four dimensional real anti-symmetrical matrixes),
\be \C&=&\left(\begin{array}{ccccc} 0 & 1
\\ -1 & 0
\end{array}\right)K\\
\T&=&\left(\begin{array}{ccccc} 0 & 1
\\ 1 & 0
\end{array}\right)K.\ee
Obviously, now $\C^2=-1$,$\T^2=1$, hence our models will now be describing class CI topological insulators.
And, as we have demonstrated, the dimensions of real IR are twice the dimensions of complex IR,
hence in the present situation the topological invariant of the basic model $\{\u(\Bp)\}$ will be $2$.
Thus the topological invariants of the powers of $n$ models will be $2n$,
agrees with the $2\ZZ$ classification of the class CI topological insulators \cite{pt3}.

It may be worth noting that, recently there are some interests in 8 bands topological insulators \cite{zhang8}\cite{Kanedirac2}, and
our present models for class CI topological insulators are 8 bands exactly.\\

To construct the BdG Hamiltonians for Class CI topological superconductors by using $4\times 4$ matrixes,
we can take the basic model $\{\u_{\k}(\Bp)\}$ as
\be \u_{\k}(\Bp)=(t\Bp^2-\mu)+(t_1p_1^2-\mu_1)\i+(t_2p_2^2-\mu_2)\j+p_3\k.\ee
Obviously $\u_{\k}(\Bp)$ is invariant under $\k\rightarrow -\k, \Bp\rightarrow -\Bp$, $\u_{-\k}(-\Bp)=\u_{\k}(\Bp)$.
And as one can see that, model $\{\u_{\k}(\Bp)\}$ is fully gapped, when $\mu/t\neq \mu_1/t_1+\mu_2/t_2$.
Now, if the quaternions are taking in two dimensional complex IR,
one will have $\nu_3(\u_k)=\deg(\u_k)$.
$\deg(\u_k)$ is nonzero only when $t\mu>0, t_1\mu_1>0, t_2\mu_2>0$ are all satisfied at the same time,
and our model will then be in topologically nontrivial phase.

To calculate the exact value of $\deg(\u_k)$ (in the nontrivial phase), we can continuously
deform (without closing the bulk gap) the parameters of $\{\u_{\k}(\Bp)\}$ to $t=t_1=t_2=\mu_1=\mu_2=1, \mu=3$.
We denote the unit modulus quaternion $\u_k/|\u_k|$ as $\u_k/|\u_k|=\hat{\u}_k=\hat{u}_0+\hat{u}_1\i+\hat{u}_2\j+\hat{u}_3\k$.
Then we take a regular point $\hat{\u}_k=1\cdot\k$, and use the topological formula for mapping degree \cite{novikov},
\be\deg(\u_k)=\sum_{\Bp_i:\ \hat{\u}_k(\Bp_i)=\k}\sign |_{\Bp_i},\ \sign |_{\Bp_i}=\sign\det\(\frac{\partial \hat{\u}_a}{\partial p_b}\)|_{\Bp_i}\ee
where $a,b=1,2,3$. As one can easily see that, the momentums that satisfy $\hat{\u}_k(\Bp_i)=\k$
are $\Bp_1=(+1, +1, 1), \Bp_2=(-1, -1, 1), \Bp_3=(+1, -1, 1), \Bp_4=(-1, +1, 1)$.
Since $\sign |_{\Bp_3}$ and $\sign |_{\Bp_4}$ are related by exchanging the two components $p_1,p_2$ of the momentum, thus $\sign |_{\Bp_3}=-\sign |_{\Bp_4}$.
Moreover, since $\hat{\u}_k(\Bp)$ is a even function of $p_1,p_2$, hence $\sign |_{\Bp_1}=\sign |_{\Bp_2}$. Thus finally we have
\be|\deg(\u_k)|=2.\label{doubling}\ee

Now let's consider the powers of $n$ models $\{\w_{\k}(\Bp)=(\u_{\k})^n\}$.
By the anticommutativity of the quaternion algebra (\ref{quaternion}), we can see that, when we multiply all the $n$ $\u_{\k}$s,
the terms of $\i\j$ must be cancelled by the terms of $\j\i$ exactly.
Hence the terms of $\k$ in $\w_{\k}(\Bp)$ can only come from the terms of $\k$ in $\u_{\k}$.
Since $\u_{-\k}(-\Bp)=\u_{\k}(\Bp)$, we then have $\w_{\k}(\Bp)=\w_{-\k}(-\Bp)$.

Now we can rewrite the quaternion $\w_{\k}(\Bp)$ as
\be\w_{\k}(\Bp)=E(\Bp)+\Delta_x(\Bp)\i+\Delta_y(\Bp)\j+d(\Bp)\k.\ee
Since $\w_{\k}(\Bp)=\w_{-\k}(-\Bp)$, we can see that, $E(\Bp), \Delta_x(\Bp), \Delta_y(\Bp)$ are all even functions of $\Bp$,
while $d(\Bp)$ is odd function of $\Bp$ with $d(-\Bp)=-d(\Bp)$.
Just as in the situation in building models for class DIII,
we can now make a unitary transformation to the off diagonal Hamiltonian, to get
\be H_{BdG}(\Bp)=\left(\begin{array}{ccccc} E(\Bp) & -d(\Bp)\k-\Delta_x\i-\Delta_y\j
\\ d(\Bp)\k+\Delta_x\i+\Delta_y\j & -E(\Bp)
\end{array}\right).\label{bdg2}\ee
One can easily verify that this BdG Hamiltonian has the particle hole symmetry and time reversal symmetry as follows
\be \C&=&\left(\begin{array}{ccccc} 0 & \i
\\ -\i & 0
\end{array}\right)K\\
\T&=&\left(\begin{array}{ccccc} \i & 0
\\ 0 & -\i
\end{array}\right)K.\ee
By noticing that in the complex IR $\i$ is a pure imaginary matrix, one can then easily get $\C^2=-1$, $\T^2=1$.
Thus our models $\{\w_{\k}(\Bp)\}$ are describing class CI topological superconductors. Noticing that,
in the topologically nontrivial phase, our basic model $\{\u_{\k}(\Bp)\}$ has $|\deg(\u_k)|=2$,
consequently the topological invariants of the powers of $n$ models $\{\w_{\k}(\Bp)\}$ are always even numbers,
in agreement with the prediction of \cite{pt3}.

The second quantized Nambu spinor that corresponds to the BdG Hamiltonian (\ref{bdg2}) can be given as,
\be\Psi(\Bp)=(c_{\uparrow}(\Bp), c_{\downarrow}(\Bp), c^{\dagger}_{\uparrow}(-\Bp), c^{\dagger}_{\downarrow}(-\Bp))^{T}.\ee
If we take $\i=i\sigma_x, \j=i\sigma_y, \k=-i\sigma_z$, and try to rewrite the BdG Hamiltonian (\ref{bdg2}) more explicitly,
(after a simple unitary transformation to get rid of some inessential imaginary unit $i$) we have
\be H_{BdG}=\left(\begin{array}{ccccc} E(\Bp) & 0 & -d(\Bp) & \bar{\Delta}(\Bp)\\
0 & E(\Bp) & \Delta(\Bp) & d(\Bp) \\
-d(\Bp) & \bar{\Delta}(\Bp) & -E(\Bp)& 0\\
\Delta(\Bp) & d(\Bp) & 0 & -E(\Bp)
\end{array}\right),\label{ci}\ee
here $\Delta(\Bp)=\Delta_x(\Bp)+i\Delta_y(\Bp)$ is the complex pairing function.

If we compare our present models (\ref{ci}) with the models (for class CI topological superconductors) that we introduced in \cite{chern},
after an appropriate reordering of the basis, we will find that,
although the specific details are different, but
the function $d(\Bp)$ in our present models can be naturally corresponded to the function $\Delta_zp^{2l+1}_z$
in the corresponding models of \cite{chern},
these two functions are all odd pairing functions. Moreover, the pairing function $\Delta(\Bp)$
in the present paper is corresponding to the pairing function $w(p_x, p_y)$ in \cite{chern}.
Both these two functions are even, and in the situations with topological invariants $2n$,
both of them are degree $2n$ polynomials of the momentum.
The pairing function $w(p_x, p_y)$ is crucial for the models of \cite{chern}.
Indeed, as \cite{chern} has demonstrated, $w(p_x, p_y)$ determines the gapless surface states. In the topologically nontrivial phase,
the degree ($2n$) of $w(p_x, p_y)$ determines that the numbers of gapless surface states should be $2n$.
Although we did not study the surface states of our present models,
but the analog between our present $\Delta(\Bp)$ and the $w(p_x, p_y)$ lets us naturally conjecture that,
in the topologically nontrivial phases with topological invariants $2n$,
the numbers of gapless surface states of our present models are exactly $2n$,
in agreement with the bulk boundary correspondence.

Still given the Hamiltonian (\ref{bdg2}), but if now we take the quaternions
in the four dimensional real IR, we then will get a class of 8 bands models for class DIII topological insulators,
with the particle hole symmetry $\C^2=1$ and time reversal symmetry $\T^2=-1$, here
\be \C&=&\left(\begin{array}{ccccc} 0 & \k
\\ -\k & 0
\end{array}\right)K\\
\T&=&\left(\begin{array}{ccccc} \k & 0
\\ 0 & -\k
\end{array}\right)K.\ee

\section{Some Relevant Mathematics of Clifford Algebras}

~~~~~To deal with the model constructions in general spatial dimensions,
we need the general theory of Clifford algebra $C_d$ (over the field of real numbers).
$C_d$ is generated by the Clifford generators $e_i, i=1,...,d$ that satisfy
\be e^2_i=-1, e_ie_j+e_je_i=0 (i\neq j).\ee
The representation matrixes of $e_i$ are gamma matrixes $\Gamma_i$.
In the IRs of $C_d$ (including both the complex IR and the real IR)
$\Gamma_i$ can always be chosen as anti-Hermitian matrixes.
Notably, in our flexible utilizations to algebraic symbols,
we will also use $e_i$ to denote the corresponding representation matrixes $\Gamma_i$,
except for the situations that may cause misunderstandings.

We will focus on the algebraic elements of $C_d$ that can be written as the form of $\u=u_0+u_ie_i$,
where $u_0, u_i(i=1,..,d)$ are real numbers, and a summation over the repeat indices $i$ is tacit.
We define the algebraic conjugation of $\u$ as $\bar{\u}=u_0-u_ie_i$.
When we look $\u$ as the representation matrix in the corresponding IR,
then this algebraic conjugation is identical with the Hermitian conjugation.
The modulus $|\u|$ of $\u$ is $|\u|=\sqrt{\bar{\u}\u}=\sqrt{u^2_0+\sum_iu^2_i}$. Obviously, if $|\u|\neq 0$,
$\u$ will be invertible, its inverse is $\u^{-1}=\bar{\u}/|\u|^2$.

The following two simple lemmas and their obvious generalizations will be useful in our model building.
$\mathbf{Lemma\ 1}$: Giving any $\u=u_0+u_ie_i$, considering $\w=\u^n$ ($n\geq 0$),
one can easily see that, $\w$ also have the form of $\w=w_0+w_ie_i$, since the cross terms
in the multiplications are cancelled pairwise due to the anti-commutativity between different Clifford generators.
Furthermore, assuming that the elements of $C_d$ are functions of some variable $x$, considering $\u(x)=u_0(x)+u_i(x)e_i$,
then we will have $\mathbf{Lemma\ 2}$: For an arbitrarily selected $l^{th}$ generator $e_l$,
if $u_l(x)$ is an odd function of $x$, while all the other functions $u_0(x), u_i(x) (i\neq l)$ are even functions of $x$,
then we will have that, among all the components of $\w(x)=\u^n(x)=w_0(x)+w_i(x)e_i$,
only the $l^{th}$ component $w_l(x)$ is odd function of $x$,
while all the others $w_0(x), w_i(x) (i\neq l)$ are even functions of $x$.
To proof this lemma, one need to notice that, in the present situation,
$\u(x)$, hence $\w(x)$, both are invariant under $x\rightarrow -x, e_l\rightarrow -e_l$.

Since we will firstly construct models for chiral $\ZZ$ topological phases,
and since these phases can only exist in odd spatial dimensions (as one can see form the periodic table),
thus we will now focus on the Clifford algebras $C_d$ with $d=2k+1$.
In this situation, it is easy to verify that $e_1e_2...e_{2k+1}$ is commuting with all the elements of $C_{2k+1}$.
By Shur's lemma we can see that, in complex IRs $e_1e_2...e_{2k+1}$ must be a scalar matrix.
On the other hand, one can easily have $(e_1e_2...e_{2k+1})^2=(-)^{k+1}$.
Thus we can conclude that, there are two inequivalent complex IRs
for $C_{2k+1}$, these two complex IRs are distinguished by the sign of $\pm$ of $e_1e_2...e_{2k+1}=\pm i^{k+1}$.
The dimensions of these complex IRs are $2^{k}$.

As for the real IRs of $C_{2k+1}$, we have the following results.
When $d=1,3,5 (\mod 8)$, the dimensions of the real IRs of $C_d$ are twice the dimensions of the corresponding complex IRs.
These real IRs can be gotten from the corresponding complex IRs by making the following replacements,
\be 1\rightarrow \left(\begin{array}{ccccc} 1 & 0
\\ 0 & 1
\end{array}\right),\
i\rightarrow \left(\begin{array}{ccccc} 0 & 1
\\ -1 & 0
\end{array}\right).\label{i}\ee
While in the case of $d=7(\mod 8)$,
the dimensions of the real IRs of $C_{d}$ are identical with the dimensions of the complex IRs.
And at $d=3,7(\mod 8)$ there are two inequivalent real IRs of $C_{d}$,
these two representations are distinguished by the sign of $\pm$ of $e_1e_2...e_{d}=\pm 1$.

Since the representation matrixes $\Gamma_i$ are anti-Hermitian matrixes,
hence in the real IRs, $\Gamma_i$ must be antisymmetrical real matrixes.
As for the situations of complex IRs,
one can always choose the representation matrixes of $e_2, e_4,..., e_{2k}$ as antisymmetrical real matrixes,
while choose the matrixes of $e_1,e_2,...,e_{2k+1}$ as symmetrical pure imaginary matrixes.

At the last of this section, we will give a generalization to the previous three form $\Ch_3$. In fact, for any invertible matrix $G$, we can define
\be \Ch_{2k+1}(G)=\frac{k!}{(2k+1)!}\frac{1}{(2\pi i)^{k+1}}\Tr (G^{-1}dG)^{\wedge 2k+1}.\ee
One can proof that, by straightforward calculations, in the sense of de Rham cohomology, there is
\be\Ch_{2k+1}(G_1G_2)=\Ch_{2k+1}(G_1)+\Ch_{2k+1}(G_2).\ee
And as we will see that the topological invariants of chiral models in $2k+1$ dimensions
can be defined by using this $\Ch_{2k+1}(G)$.

\section{Models For All Chiral $\ZZ$ Topological Phases}

\subsection{Chiral $\ZZ$ Topological Phases In General}

~~~~~Now we will use the basic ideas that we have developed in the three dimensional situation,
to construct models for all chiral
$\ZZ$ topological insulators in odd dimensions.
In our constructions the general models $\{\w(\Bp)\}$ with topological invariants $n$ are constructed
as the powers of
$n$ models of some basic model $\{\u(\Bp)\}$.

Now, we take $\u(\Bp)=u_0(\Bp)+u_i(\Bp)\Gamma_i$,
the summation over $i$ is tacit. According to the well known proposal\cite{pt2,pt3}
for the topological invariants of chiral models, we can calculate the topological invariants
$\nu_{2k+1}(\u^n)$ of $\{\u^n(\Bp)\}$ ($|\u(\Bp)|\neq 0$) as follows (by using some related mathematical facts
that we summarized in last section)
\be \nu_{2k+1}(\u^n)&=&\int_{\Bp} \Ch_{2k+1}(\u^n)=n\int_{\Bp}\Ch_{2k+1}(\u)\\
&=&\pm \frac{n}{\mathrm{Vol}(S^{2k+1})(2k+1)!}\int_{\Bp}\epsilon^{i_0i_1...i_{2k+1}}
\frac{u_{i_0}du_{i_1}\wedge...\wedge du_{i_{2k+1}}}{|\u|^{2k+2}}\label{ch21}\\
&=&\pm n\deg(\u),\label{ch2}\ee
where $\deg(\u)$ denotes the mapping degrees of $\Bp\rightarrow\u(\Bp)/|\u|$,
or equivalently of $\Bp\rightarrow (u_0/|\u|,u_1/|\u|,...,u_{2k+1}/|\u|)$.
This map maps the $d=2k+1$ dimensional momentum space to the unit sphere $S^{2k+1}$ of $\BR^{2k+2}$.

To get (\ref{ch21}), we need to trace over $\Gamma_1\Gamma_2...\Gamma_{2k+1}$.
And since we have used $e_1e_2...e_{2k+1}=\pm i^{k+1}$ and have used the dimensions $2^k$ of the complex IRs of $C_{2k+1}$,
the final result (\ref{ch2}) can only be applied in the situations of complex IRs.
If what we are considering is the real IR,
then for the case of $d=7 (\mod 8)$, we will get the same $\pm n\deg(\u)$,
since in this situation the dimensions of real IR is also $2^k$.
But for the case of $d=3 (\mod 8)$, we will get $\pm 2n\deg(\u)$, since
in this situation the dimensions of real IR is twice the dimensions of the corresponding complex IR.

Moreover, for the cases of $d=1, 5(\mod 8)$(in the real IR),
instead of (\ref{ch2}) we will get a $0$. This can be understood as follows.
Firstly, in the complex IR of $d=1, 5(\mod 8)$, we have $e_1e_2...e_{d}=\pm i$.
Then, when we turn to the real IR, we should make the replacement (\ref{i}).
And, when we apply this replacement to the right hand side of
$e_1e_2...e_{d}=\pm i$, we will get a traceless matrix. Finally, to get the final result (\ref{ch2}),
we need to trace over $\Gamma_1\Gamma_2...\Gamma_{d}$, this is now a traceless matrix, thus resulting $0$.

If fact, for any model $\{\w(\Bp)\}$, with $\w(\Bp)$ given by the algebraic element
$\w(\Bp)=w_0(\Bp)+w_i(\Bp)e_i$.
The likewise calculation tells us
\be\nu_{2k+1}(\w)=\int_{\Bp}\Ch_{2k+1}(\w)=\deg(\w).\label{ti}\ee
Thus, the topological invariants of this kind of models are given by the mapping degree of
the map $\Bp\rightarrow\w/|\w|$ from the $2k+1$ dimensional momentum space to $S^{2k+1}$.
On the other hand, the Hopf theorem\cite{novikov} tells us, all maps from momentum space to $S^{2k+1}$ that have the same mapping degree
are homotopically equivalent. Thus, any model with a given $\deg(\w)$ must
be homotopically equivalent to some powers of $n$ model.
Hence, the constructions of the present paper can be
regarded as a concrete realization for topological phases within any homotopical classes.

\subsection{The First Type of Constructions}

~~~~~Now we can give a specific construction for the basic model $\{\u(\Bp)=u_0(\Bp)+u_i(\Bp)e_i\}$, with
\be (u_0(\Bp),u_1(\Bp),...,u_{2k+1}(\Bp))=(m+t\Bp^2,p_1,...,p_{2k+1}).\ee
As one can easily see that, when $\Bp\rightarrow 0$, $\frac{\u}{|\u|}(\Bp)\rightarrow \sign(m)$,
while when $|\Bp|\rightarrow \infty$, $\frac{\u}{|\u|}(\Bp)\rightarrow \sign(t)$.
Thus, if $mt>0$, the map $\Bp\rightarrow \u/|\u|$ will cover only half of the sphere $S^{2k+1}$, hence $\deg(\u)=0$.
While if $mt<0$, then the map $\Bp\rightarrow \u(\Bp)/|\u|$ will cover the whole $S^{2k+1}$ exactly once, thus $\deg(\u)=1\cdot \sign(m)$.
That is to say, when $mt<0$, our basic model $\{\u(\Bp)\}$ will be in topologically nontrivial phase with minimum nonzero topological invariant.

We can then construct models with arbitrary topological invariants by an appropriate powers of $n$ model $\{\w(\Bp)=\u^n(\Bp)\}$.
By the lemma 1 of last section, we can see that $\w(\Bp)$ can be written as
\be \w(\Bp)=E(\Bp)+d_i(\Bp)e_i.\label{wed}\ee
Noticing that $u_0(\Bp)$ is an even function of $\Bp$, while $u_i(\Bp),i=1,..,2k+1$ are all odd functions of $\Bp$,
hence by the lemma 2, we can know that,
$E(\Bp)$ must be an even function, while $d_i(\Bp)$ must be odd functions.

By using (\ref{wed}), we can rewrite our models $\{\w(\Bp)=\u^n(\Bp)\}$ as
\be \left(\begin{array}{ccccc} 0 & E(\Bp)+d_i(\Bp)\Gamma_i
\\ E(\Bp)-d_i(\Bp)\Gamma_i & 0
\end{array}\right).\label{h1}\ee
By a simple unitary transformation, we can rewrite (\ref{h1}) as a more usual form,
\be \left(\begin{array}{ccccc} E(\Bp) & d_i(\Bp)\Gamma^{\dagger}_i
\\ d_i(\Bp)\Gamma_i & -E(\Bp)
\end{array}\right).\label{h11}\ee
In what follows, we will try to identify the discrete symmetries of this kind of models (\ref{h1}).

We will firstly discuss the situations of complex IRs.
Hence at the dimensions of $d=3,7(\mod 8)$,
we have the charge conjugation matrix $C$, with
\be C\Gamma_i C^{-1}=-\Gamma^{T}_i=\Gamma^{*}_i.\ee
$C$ can be explicitly chosen as $C=\Gamma_2\Gamma_4...\Gamma_{2k}$, it is of course a real matrix.
One can then easily get that, for the cases of $d=3 (\mod 8)$, $C^2=-1$,
while for $d=7 (\mod 8)$, $C^2=1$.
At these two kinds of spatial dimensions,
we can find the particle hole symmetry $\C$ and time reversal symmetry $\T$ of our
models (\ref{h1}), they act as (in the off diagonal basis)
\be \C&=&\left(\begin{array}{ccccc} 0 & C
\\ -C & 0
\end{array}\right)K,\\
\T&=&\left(\begin{array}{ccccc} 0 & C
\\ C & 0
\end{array}\right)K.\ee
Obviously, at $d=3 (\mod 8)$ dimensions, we have $\C^2=1,\T^2=-1$.
Thus our models are describing the class DIII topological insulators.
While, at $d=7 (\mod 8)$, we have $\C^2=-1, \T^2=1$,
thus our models are describing the class CI topological insulators.

Still at $d=3,7(\mod 8)$ dimensions,
but if our Clifford algebras are now taking real IRs,
then the particle hole symmetry and the time reversal symmetry will be
(in the off diagonal basis)
\be \C&=&\left(\begin{array}{ccccc} 0 & 1
\\ -1 & 0
\end{array}\right)K,\\
\T&=&\left(\begin{array}{ccccc} 0 & 1
\\ 1 & 0
\end{array}\right)K. \ee
Obviously, now we always have $\C^2=-1$, $\T^2=1$.
Consequently, in this situation,
our models ($\ref{h1}$) always describe class CI topological insulators (both
at $d=3(\mod 8)$ dimensions and at $d=7(\mod 8)$ dimensions).
Fortunately, at the dimensions of $d=3 (\mod 8)$,
the dimensions of real IRs are twice the dimensions of complex IRs,
hence the topological invariants of our constructions will be even numbers,
agrees with the $2\ZZ$ classification of $d=3 (\mod 8)$ class CI topological insulators \cite{pt3}.
While at $d=7 (\mod 8)$, the dimensions of real IRs are identical with the complex IRs,
thus in this case, the topological invariants of our models can be arbitrary integers,
also in agreement with the classification in the periodic table.

Still taking the complex IR, but if we now consider the $d=1,5(\mod 8)$ situations.
In these cases, we still have a charge conjugation matrix $C$, but now $C$ acts as
\be C\Gamma_iC^{-1}=\Gamma^{T}_i=-\Gamma^{*}_i.\ee
Here $C$ can be chosen as $C=\Gamma_1\Gamma_3...\Gamma_{2k+1}$ explicitly, it is of course a pure imaginary matrix now.
Obviously, at $d=1(\mod 8)$, $C^2=-1$, while at $d=5(\mod 8)$, $C^2=1$.
At these two kinds of spatial dimensions,
the particle hole symmetry and time reversal symmetry of ($\ref{h1}$) can be given as
\be \C&=&\left(\begin{array}{ccccc} C & 0
\\ 0 & -C
\end{array}\right)K\\
\T&=&\left(\begin{array}{ccccc} C & 0
\\ 0 & C
\end{array}\right)K.\ee
Thus, at the dimensions of $d=1(\mod 8)$, we have $\C^2=\T^2=1$, hence our models (\ref{h1})
are now describing class BDI topological insulators.
While at the dimensions of $d=5(\mod 8)$, we have $\C^2=\T^2=-1$, our models
(\ref{h1}) will then describe the class CII topological insulators.
And as we have seen, at these two kinds of dimensions,
there are no topologically nontrivial phases when the Clifford algebras are taking real IRs.

\subsection{The Second Type of Constructions}

~~~~~We now try to give the second type of specific constructions, by taking the basic model $\{\u(\Bp)=u_0(\Bp)+u_i(\Bp)e_i\}$ as
\be (u_0,u_1,...,u_{2k+1})=(m+t\Bp^2,p_1, m_2+t_2p_2^2, p_3,..,m_{2k}+t_{2k}p_{2k}^2,p_{2k+1}).\label{construction2}\ee
Noticing that $u_0,u_2,...,u_{2k}$ are even functions of $\Bp$, while $u_1,u_3,...,u_{2k+1}$ are odd functions.
Thus the mapping degree $\deg(\u)$ of map $\Bp\rightarrow \u/|\u|(\Bp)$
can be nonzero only when $d=1, 5(\mod 8)$.
The reason is that, for the situations of $d=3, 7(\mod 8)$,
we will have even numbers of odd functions among the various components of $\u(\Bp)$,
hence the values of these odd functions at $-\Bp$ always can be rotated back to the values at $\Bp$,
consequently the map $\Bp\rightarrow \u/|\u|$ can not cover the whole $S^{2k+1}$,
thus at the dimensions of $d=3, 7(\mod 8)$ we must have $\deg(\u)=0$.
Hence, for the present type of constructions (\ref{construction2}),
we only need to focus on the models at the dimensions of $d=1, 5(\mod 8)$.

As one can see that, $\deg(\u)$ will be nonzero only if $mt<0, m_2t_2<0 ,...,m_{2k}t_{2k}<0$ are all satisfied at the same time.
For the situations of $d=1(\mod 8)$, through a likewise discussion as that in last subsection, we can know
\be \deg(\u)=\sign(m)\sign(m_2)...\sign(m_{2k}).\ee
While, for the situations of $d=5(\mod 8)$, detail analysis tells us
\be \deg(\u)=2\cdot\sign(m)\sign(m_2)...\sign(m_{2k}).\ee
Here the mathematical mechanism for the doubling of the mapping degree is
just like the mechanism that lead to (\ref{doubling}), which we have discussed in details.

Just as before, the models for topological insulators with arbitrary topological invariants
can be constructed as the powers of $n$ models $\{\w(\Bp)=\u^n(\Bp)\}$.
And we can always write $\w(\Bp)$ as
\be\w(\Bp)=E(\Bp)+D_{2i}(\Bp)e_{2i}+d_{2j-1}(\Bp)e_{2j-1},\ee
summations over the even indices
and over the odd indices respectively are tacit.
Since the odd components $u_{2j-1}$ of $\u$ are all odd functions of $\Bp$,
while the other components are all even functions, by lemma 2 we can see that,
$E(\Bp), D_{2i}(\Bp)$ are even functions of $\Bp$,
while $d_{2j-1}(\Bp)$ are odd functions of $\Bp$.

As we have known, at the dimensions of $d=1, 5(\mod 8)$,
only the complex IRs of the Clifford algebras need to be considered.
In these representations, the matrixes of $e_{2i}$ are real, and the matrixes of $e_{2j-1}$ are purely imaginary.
Combing the parities of the functions $E(\Bp), D_{2i}(\Bp)$ and $d_{2j-1}(\Bp)$,
we can find that the particle hole symmetry and the time reversal symmetry of our models $\{\w(\Bp)\}$ are
\be \C&=&\left(\begin{array}{ccccc} 1 & 0
\\ 0 & -1
\end{array}\right)K\\
\T&=&K.\ee
Obviously, $\C^2=\T^2=1$, hence our models
are describing the class BDI topological insulators.

The subtlety is, at the dimensions of $d=1(\mod 8)$,
our basic model $\{\u(\Bp)\}$ will have $|\deg(\u)|=1$ in the topologically nontrivial phase,
hence in this situation our general constructions for class BDI topological insulators
can have any integer topological invariants. While at the dimensions of $d=5(\mod 8)$,
since the basic model $\{\u(\Bp)\}$ have $|\deg(\u)|=2$,
hence the general models can only have the topological invariants of even numbers,
fully agrees with the predictions\cite{pt3} of the periodic table.\\

Now we slightly alter our previous construction for the basic model $\{\u(\Bp)\}$ to take
\be (u_0,u_1,...,u_{2k+1})=(m+t\Bp^2,m_1+t_1p^2_1, p_2,..,p_{2k},m_{2k+1}+t_{2k+1}p^2_{2k+1}),\ee
i.e. $u_0=m+t\Bp^2$,$u_{2i}=p_{2i}$,$u_{2j-1}=m_{2j-1}+t_{2j-1}p^2_{2j-1}$.
The likewise analysis tells us, the mapping degree $\deg(\u)$ can be nonzero only at the dimensions of $d=3,7(\mod 8)$,
thus we will now focus on these two kinds of spatial dimensions. And,
we will focus on the situations of complex IRs of the Clifford algebras.

Obviously, the mapping degree $\deg(\u)$ can be nonzero, only when $mt<0, m_1t_1<0,...,m_{2k+1}t_{2k+1}<0$ are all satisfied.
Detail calculations show us, at the dimensions of $d=3(\mod 8)$,
\be\deg(\u)=2\cdot\sign(m)\sign(m_1)\sign(m_3)...\sign(m_{2k+1}).\ee
While at the dimensions of $d=7(\mod 8)$ the mapping degrees are
\be\deg(\u)=\sign(m)\sign(m_1)\sign(m_3)...\sign(m_{2k+1}).\ee

Just as before, we can construct $\{\w(\Bp)=\u^n(\Bp)\}$. Here $\w(\Bp)$ can be written as
\be\w(\Bp)=E(\Bp)+D_{2j-1}(\Bp)e_{2j-1}+d_{2i}(\Bp)e_{2i}.\ee
By utilizing lemma 2, we can easily see that $E(\Bp),D_{2j-1}(\Bp)$ are even functions of $\Bp$,
while $d_{2i}(\Bp)$ are odd functions of $\Bp$.
Noticing that $e_{2j-1}$ are pure imaginary matrixes, while
$e_{2i}$ are real matrixes, one can then easily find the $\C$ symmetry and the $\T$ symmetry of our present constructions,
\be \C&=&\left(\begin{array}{ccccc} 0 & 1
\\ -1 & 0
\end{array}\right)K\\
\T&=&\left(\begin{array}{ccccc} 0 & 1
\\ 1 & 0
\end{array}\right)K.\ee
Obviously $\C^2=-1$,$\T^2=1$, thus our present constructions are
describing the $d=3,7(\mod 8)$ class CI topological insulators.
And the previous calculations to the mapping degree $\deg(\u)$ tell us that,
at the dimensions of $d=3(\mod 8)$, our models can realize any even topological invariants,
while at the dimensions of $d=7(\mod 8)$ our models can realize any integer topological invariants.
This is of course in agreement with the classifications of \cite{pt3}.

\subsection{The $d=7(\mod 8)$ Class $\mathrm{DIII}$ and $d=1(\mod 8)$ Class $\mathrm{CII}$}

~~~~~As one may noticed, so far we have constructed models for all the chiral $\ZZ$ topological phases,
except for the $d=7(\mod 8)$ Class $\mathrm{DIII}$ and the $d=1(\mod 8)$ Class $\mathrm{CII}$.
In this subsection, we will try to give the constructions for these two symmetry classes.

We will firstly discuss the $d=7(\mod 8)$ Class $\mathrm{DIII}$ topological phases.
To give the corresponding model constructions,
let's consider the real IR of the Clifford algebra $C_d$, with $d=7(\mod 8)$.
In this representation, all the Clifford generators $e_i$ are represented as real anti-symmetrical matrixes.

We now take the algebraic element $\u(\Bp)=u_0(\Bp)+u_i(\Bp)e_i$ of the basic model $\{\u(\Bp)\}$ as
\be (u_0,u_1,...,u_{d})=(m+t\Bp^2,m_1+t_1p^2_1, m_2+t_2p^2_2,..,m_{d-1}+t_{d-1}p^2_{d-1},p_d),\ee
i.e. only the last component $u_d$ is an odd function of the momentum, with $u_d=p_d$. Likewise,
the mapping degree $\deg(\u)$ can be nonzero, only when $mt<0, m_1t_1<0,...,m_{d-1}t_{d-1}<0$ are all satisfied at the same time.
In this situation, detail calculations show us
\be\deg(\u)=2\cdot\sign(m)\sign(m_1)...\sign(m_{d-1}).\ee
Pay attention to the factor of $2$, one may want to know in what situations the degrees of mapping will be doubled.
The general rule is,
if among all the components of $\u(\Bp)$ (except $u_0$), there are $2 (\mod 4)$
components are taking the form of $u_i=m_i+t_ip^2_i$,
then the mapping degree will be doubled.
The mathematical mechanism behind this rule has been explained in the reasons lead to
(\ref{doubling}).

We now take $\{\w(\Bp)=\u^n(\Bp)\}$, here $\w(\Bp)$ can be written as
\be \w(\Bp)=E(\Bp)+\sum^{d-1}_iD_i(\Bp)e_i+d(\Bp)e_d.\ee
Where, as one can easily proof by using the lemma 2, $E(\Bp), D_i(\Bp)$ are even functions of $\Bp$,
while $d(\Bp)$ is an odd function.
Noticing that we are taking the real IR,
one can then easily find the $\C$ symmetry and the $\T$ symmetry of $\{\w(\Bp)\}$, they are
\be\C&=&\left(\begin{array}{ccccc} 0 & e_d
\\ -e_d & 0
\end{array}\right)K\\
\T&=&\left(\begin{array}{ccccc} 0 & e_d
\\ e_d & 0
\end{array}\right)K.\ee
Obviously $\C^2=1$, $\T^2=-1$, thus our models are indeed describing the $d=7(\mod 8)$
Class $\mathrm{DIII}$ topological insulators (superconductors).

It is worth noting that, since $|\deg(\u)|=2$, hence the topological invariants of our models $\{\w(\Bp)\}$
are classified by $2\ZZ$. Moreover,
since for the situations of $d=7(\mod 8)$ the dimensions of the real IR is identical
with the dimensions of the complex IR, there are no dimensional doubling,
hence we have no additional factor of 2.
Thus, our constructions are classified just by $2\ZZ$, not by $4\ZZ$,
fully agrees with the prediction \cite{pt3} of the periodic table.\\

To construct models for $d=1(\mod 8)$ Class $\mathrm{CII}$ topological insulators,
let's consider the complex IRs of $d=1(\mod 8)$ Clifford algebras.
After some tries and failures,
one will find that all the constructions that we have employed up to now are failed to realize
the symmetries (i.e. $\C^2=\T^2=-1$) of $\mathrm{CII}$ symmetry class at $d=1 (\mod 8)$.
To construct models with this kind of symmetries, one way is to take the $\u(\Bp)$ of our basic model $\{\u(\Bp)\}$
as \be\u(\Bp)=\v(\Bp)\otimes (a'+b'\i+c'\k),\label{1cons}\ee
where $a',b',c'$ are nonzero real constants,
and $\i,\k$ are two generators of the quaternion,
and these quaternions are taking in two dimensional complex IR (In this representation $\i,\k$ are pure imaginary
$2\times 2$ matrixes). The $\v(\Bp)$ in (\ref{1cons})
is an element of the $d=1(\mod 8)$ Clifford algebra, with $\v(\Bp)=v_0(\Bp)+v_i(\Bp)e_i$.
Moreover, the components of $\v(\Bp)$ can be taken as
\be (v_0(\Bp),v_1(\Bp),...,v_{d}(\Bp))=(m+t\Bp^2,p_1,...,p_{d}).\ee
If we now calculate the topological invariant $\int_{\Bp} \Ch_{d}(\u)$, we will get
(since the numbers of bands of our models are doubled by the quaternions),
\be\nu_d(\u)=\int_{\Bp} \Ch_{d}(\u)=2\cdot\deg(\v).\ee
Obviously, $\deg(\v)$ can be nonzero only when $mt<0$, and if this is satisfied,
we will have $\deg(\v)=\sign(m)$.

As we now are quite familiar, the general models can be constructed as $\{\w(\Bp)=\u^n(\Bp)\}$.
Where $\w(\Bp)$ can be written as
\be\w(\Bp)=(E(\Bp)+d_i(\Bp)e_i)\otimes (a+b\i+c\k),\ee
where $a,b,c$ are real constants. By using the lemma 2 we can easily see that $E(\Bp)$ is an even function of $\Bp$,
while $d_i(\Bp)$ are odd functions.

Now one can easily find the $\C$ symmetry and the $\T$ symmetry of $\{\w(\Bp)\}$, given by
\be \C&=&\left(\begin{array}{ccccc} C & 0
\\ 0 & -C
\end{array}\right)\otimes \j\cdot K\\
\T&=&\left(\begin{array}{ccccc} C & 0
\\ 0 & C
\end{array}\right)\otimes\j\cdot K.\ee
Here $C$ is the charge conjugation matrix in the $d=1(\mod 8)$ Clifford algebra,
it is a pure imaginary matrix, satisfying $C\Gamma_iC^{-1}=\Gamma^{T}_i=-\Gamma^{*}_i$, and $C^2=-1$.
We can easily see that $\C^2=\T^2=-1$,
thus our constructions are indeed describing the $d=1(\mod 8)$ Class $\mathrm{CII}$ topological insulator.
And since in the topologically nontrivial phase our basic model $\{\u\}$ has $|\nu_d(\u)|=2$,
our present general constructions are consequently classified by $2\ZZ$, in agreement with the classification of \cite{pt3}.

This complete our model constructions for all the chiral $\ZZ$ topological phases.

\subsection{Bott Periodicity}

~~~~~As one have seen, in all the constructions that we have studied,
the spatial dimensions always appear with period mod 8.
This is of course the well known Bott periodicity of the periodic table
that are mostly naturally explained in terms of K theory \cite{pt21}.
In this subsection we will brief discuss,
from the point of view of model constructions,
the connections between this Bott periodicity of topological phases
and the Bott periodicity of the Clifford algebras.

The Bott periodicity of Clifford algebras tells us,
\be C_d\otimes C_8\simeq C_{d+8},\label{bottp}\ee
here $C_8$ is the Clifford algebra at the dimensions of 8, with Clifford generators
$e_{d+i}, i=1,...,8$. Since any IR of $C_8$ is isomorphic to the real IR, which has the dimensions of 16,
thus we have $C_8\simeq \BR(16)$, here $\BR(16)$ stands for the algebra of $16\times 16$ real matrixes.
We will see that this means that, in any symmetry classes,
the numbers of bands of the models at the dimensions of $d+8$
are 16 times as that of the models at dimensions $d$.
If we denote the Clifford generators of $C_{d+8}$ as $\tilde{e}_i, i=1,...,d+8$,
then (\ref{bottp}) can be explicitly realized as
\be \tilde{e}_i=e_i\otimes({e}_{d+1}{e}_{d+2}...{e}_{d+8})(i=1,...,d), \tilde{e}_{d+i}=1\otimes {e}_{d+i}(i=1,...,8).\label{bottpp}\ee

This Bott periodicity of the Clifford algebras means that for each $d$ dimensional model
$\{\w(\Bp)\}$, with $\w(\Bp)=w_0(\Bp)+\sum^{d}_{i=1}w_i(\Bp)e_i$,
we can associate a $d+8$ dimensional model $\{\tilde{\w}(\Bp)=\tilde{w}_0(\Bp)+\sum^{d+8}_{i=1}\tilde{w}_i(\Bp)\tilde{e}_i\}$,
with $\tilde{e}_i$ given by (\ref{bottpp}) and the components of $\tilde{\w}(\Bp)$ given by
\be \tilde{w}_i(\Bp)=w_i(\Bp)(i=0,...,d),\ \tilde{w}_i(\Bp)=p_i(i=d+1,...,d+8).\label{bottppp}\ee
This is the Bott periodicity of our model constructions.

On the other hand, as we have seen that, there are various phenomena of period mod 4 in our previous model constructions.
In terms of Clifford algebras, this period mod 4 corresponds to the relationship $C_d\otimes C_4\simeq C_{d+4}$.
But Clifford algebras are not period mod 4, since $C_4\simeq \BH(2)$
($\BH(2)$ stands for the algebra of $2\times 2$ matrixes over quaternion)
is not a matrix algebra over real numbers.
Likewise, our model constructions are not period mod 4 too (but period mod 8). This can be seen from
the detail constructions that we have studied, in these constructions, the models at $d+4$ dimensions
either have different $\C, \T$ symmetries,
or have different topological classifications
(by $\ZZ$ or by $2\ZZ$) with the models at $d$ dimensions.
Thus they can not be related by some isomorphism like (\ref{bottppp}).

\section{Model Constructions in Even Dimensions}

~~~~~Having complete the model constructions for all $\ZZ$ topological phases in odd dimensions $d=2k+1$.
We now turn to the model constructions for all $\ZZ$ topological phases in even dimensions $d=2k+2$.
To achieve this, we will use the well known connection between models at $d=2k+2$ and models at $d=2k+1$ \cite{pt3}.

Indeed, for any $2k+1$ dimensional model $\{\w(\Bp)\}$ that we have constructed in last section,
we can always construct a dimensional lifted $2k+2$ dimensional model
\be H_{2k+2}=\left(\begin{array}{ccccc} p_{2k+2} & \w'(\Bp, p_{2k+2})
\\ \bar{\w}'(\Bp, p_{2k+2}) & -p_{2k+2}
\end{array}\right).\label{2k+2}\ee
Where $\w'(\Bp, p_{2k+2})$ is an element of the Clifford algebra $C_{2k+1}$, this element is gotten
by extending the Clifford element $\w(\Bp)$ from the $2k+1$ dimensional momentum space to the $2k+2$ dimensional momentum space,
with \be\w'(\Bp, 0)=\w(\Bp).\ee

As demonstrated by \cite{pt3}, the topological invariant (i.e. the ${k+1}^{th}$ Chern numbers $c_{k+1}$)
of this $2k+2$ dimensional model (\ref{2k+2}) is equal to the winding number $\nu_{2k+1}$
of the related $2k+1$ dimensional model $\{\w(\Bp)\}$
\be c_{k+1}(H_{2k+2})=\nu_{2k+1}(\w)=\int_{\Bp}\Ch_{2k+1}(\w).\ee
Since we have constructed $2k+1$ dimensional models with any topological invariants,
we can thus get $2k+2$ dimensional models with any topological invariants.

To see these $2k+2$ models more clearly, we now rewrite the Clifford element
$\w(\Bp)$ as $\w(\Bp)=E(\Bp)+d_i(\Bp)e_i$, here $d_i(\Bp)$ need not to be odd functions.
And we also rewrite the corresponding $\w'(\Bp, p_{2k+2})$ as $\w'(\Bp, p_{2k+2})=E'(\Bp, p_{2k+2})+d'_i(\Bp, p_{2k+2})e_i$.
Hence our $2k+2$ dimensional model (\ref{2k+2}) can be written as
\be \left(\begin{array}{ccccc} p_{2k+2} & E'(\Bp, p_{2k+2})+d'_i(\Bp, p_{2k+2})e_i
\\ E'(\Bp, p_{2k+2})+d'_i(\Bp, p_{2k+2})\bar{e}_i & -p_{2k+2}
\end{array}\right).\label{evenh}\ee
Naturally, we can define some new Clifford generators $e'_i (i=1,...,2k+1)$, $e'_{2k+2}$, and $e'$, by
\be e'&=&\left(\begin{array}{ccccc} 0 & 1
\\ 1 & 0
\end{array}\right), \label{90}\\
e'_{2k+2}&=&\left(\begin{array}{ccccc} 1 & 0
\\ 0 & -1
\end{array}\right),\label{91}\\
e'_i&=&\left(\begin{array}{ccccc} 0 & e_i
\\ \bar{e}_i & 0
\end{array}\right).\label{92}\ee
By using these new Clifford generators we can rewrite our $2k+2$ dimensional model as,
\be H_{2k+2}&=&E'(\Bp, p_{2k+2})e'+\sum^{2k+2}_{i=1}d'_i(\Bp, p_{2k+2})e'_i,\\
&=&m'(\Bp, p_{2k+2})\Gamma'+\sum^{2k+2}_{i=1}d'_i(\Bp, p_{2k+2})\Gamma'_i.\label{gammaa}\ee
Where we have set $d'_{2k+2}(\Bp, p_{2k+2})=p_{2k+2}$,
and we have rewritten $E'(\Bp, p_{2k+2})$ as $m'(\Bp, p_{2k+2})$,
since it is the mass term of the $2k+2$ dimensional model. The gamma matrixes in (\ref{gammaa}) are
just the representation matrixes of their corresponding Clifford generators.

One can easily verify that these new Clifford generators $e'_i(i=1,...,2k+2)$ and $e'$ satisfy $e'^2=e'^2_i=1$,
$e'_ie'_j+e'_je'_i=0(i\neq j)$, $e'e'_i+e'_ie'=0$. Hence they generate the $2k+3$ dimensional Clifford algebra $C'_{2k+3}$.
This kind of Clifford algebras are different with the previous $C_{2k+1}$,
since their generators are square to $+1$ rather than to $-1$.
We will denote this kind of Clifford algebras as $C'_{p}$ (with $p$ generators).
As we have seen that, while the Clifford algebras behind the model building for chiral $\ZZ$ topological phases are $C_d$,
the Clifford algebras behind the model building for $\ZZ$ topological phases in even dimensions are $C'_{d+1}$.

In terms of the Clifford algebras, the connection between the dimensional lifted
$2k+2$ dimensional models and the $2k+1$ dimensional models are
just the well known
\be C_{2k+1}\otimes C'_2\simeq C'_{2k+3}.\ee
This isomorphism between Clifford algebras is realized by (\ref{90}),(\ref{91}),(\ref{92}).

The discrete symmetries ($\C$ or $\T$) of our $2k+2$ dimensional models are different with
but are related to the symmetries of the original $2k+1$ dimensional models.
Firstly, as one can see that, the dimensional lifting (\ref{2k+2})
breaks the chiral symmetry $S$ of the original $2k+1$ dimensional models.
Thus our $2k+2$ dimensional models can have either the $\C$ symmetry or the $\T$ symmetry, but not the both.
In what follows, we will take the first type of constructions (discussed in section (4.2)) as examples,
to illustrate what kind of symmetry can be retained in their $2k+2$ dimensional lifting.

In the first type of constructions,
the function $E(\Bp)$ is even, while the functions $d_i(\Bp)$ are odd.
Thus we further require that the $2k+2$ dimensional extended functions $E'(\Bp, p_{2k+2})$ and $d'_i(\Bp, p_{2k+2})$
must have the same parities with the corresponding $E(\Bp)$ and $d_i(\Bp)$ respectively.
For the situations of complex IRs, it is now easy to see that, at the dimensions of $d=3+1, 7+1(\mod 8)$
the dimensional lifted models (\ref{evenh}) invalid the $\C$ symmetry, but the $\T$ symmetry is maintained.
Since the $d=3 (\mod 8)$ models are in the symmetry class DIII (hence $\T^2=-1$),
thus the $d=4 (\mod 8)$ lifted models are in symmetry class AII ($\T^2=-1$).
Likewise, since the $d=7 (\mod 8)$ models are in class CI (hence $\T^2=1$),
thus our $d=8 (\mod 8)$ models are in class AI ($\T^2=1$).

Also at the dimensions of $d=3+1, 7+1(\mod 8)$, but if now the Clifford algebras are taking in real IRs,
in these situations our dimensional lifted models (\ref{evenh}) will also invalid the $\C$
symmetry, and maintain the $\T$ symmetry.
Since the $d=3, 7 (\mod 8)$ dimensional models are always in class CI,
hence the corresponding $d=4, 8(\mod 8)$ dimensional models are always in class
AI(with $\T^2=1$).

If what we are considering are the dimensional lifted models at the dimensions of $d=1+1, 5+1(\mod 8)$,
then we will need only to consider the complex IRs.
One can easily see that, in these cases, the $\T$ symmetry of the $d=1, 5(\mod 8)$ models is invalided,
but the $\C$ symmetry is maintained. Since the $d=1(\mod 8)$ models are in class BDI (hence $\C^2=1$),
thus the lifted $d=2(\mod 8)$ models are in class D ($\C^2=1$).
Likewise, since the $d=5(\mod 8)$ models are in class CII (hence $\C^2=-1$),
thus the $d=6(\mod 8)$ lifted models are in class C ($\C^2=-1$).

As one can check that, these kinds of relationships between the discrete symmetries of the $2k+1$ dimensional models and the symmetry of
the $2k+2$ dimensional lifted models can be generalized to all the constructions that we have studied in last section.
The results can be summarised as follows, as one lifts a $2k+1$ dimensional model to a model at $2k+2$ dimensions,
the symmetry classes of these models are shifted according
\be\mathrm{BDI}\rightarrow\mathrm{D},\mathrm{DIII}\rightarrow\mathrm{AII},
\mathrm{CII}\rightarrow\mathrm{C},\mathrm{CI}\rightarrow\mathrm{AI}.\ee
This relationship is firstly discussed in \cite{pt3} (with opposite arrows).

Thus we complete the model constructions for all the $\ZZ$ topological phases of the eight real symmetry classes.

\section{Models for $\ZZ_2$ Topological Insulators}

~~~~In this section, we will turn to the model constructions
for all $\ZZ_2$ topological phases. Especially,
we will give a \emph{theoretical derivation} for the well known model \cite{TI}
for $\mathrm{Bi_2Se_3}, \mathrm{Bi_2Te_3}$ and $\mathrm{Sb_2Te_3}$
and for the BHZ model for quantum spin Hall insulator.

The basic idea that we will employ is that any $\ZZ_2$ topological insulators (superconductors) at $d-1$ dimensions or at $d-2$ dimensions
can be realized as the descendants of some parent $\ZZ$ topological insulators (in the same symmetry class) at $d$ dimensions,
through the mechanism of dimensional reduction. This mechanism was first put forward by\cite{pt1} to
connect the two dimensional quantum spin Hall insulator and
the three dimensional time reversal invariant topological insulator to
the four dimensional time reversal invariant Chern insulator.
\cite{pt1} also use this idea to give a general $\ZZ_2$ classifications
for the time reversal invariant topological insulators in two and three dimensions.
And this has been generalized to all $\ZZ_2$ topological phases by \cite{pt3}.

In this dimensional reduction, the $\ZZ_2$ classification of the descendent topological insulators
is determined by the parity of the topological invariant of the parent $\ZZ$ topological insulator.
For this reason, the topological insulators that are classified by $2\ZZ$ rather than by $\ZZ$
(such as, the class CII topological insulators in 1 dimension, the class C topological superconductors in 2 dimensions,
and the class CI topological superconductors in 3 dimensions, etc.)
will have no nontrivial $\ZZ_2$ descendants.

Conversely, the absence of nontrivial $\ZZ_2$ topological phases in some $d$ dimensions
(such as the symmetry class CII at 0 dimension, the class C at 1 dimension, and the class CI at 2 dimensions, etc.)
will mean that the corresponding $d+1$ dimensional topological insulators must
be either trivial or be classified by $\ZZ_2$ or by $2\ZZ$, but can not be classified by $\ZZ$.
In fact this is a general pattern of the periodic table, and as we have mentioned repeatly, our model constructions are exactly in
accordance with this pattern.

Now we can construct the models for $\ZZ_2$ topological phases as follows.
Assuming that we have a model for some $d+1$ dimensional $\ZZ$ topological phase,
with the Bloch Hamiltonian $H_{d+1}$. To construct the $d$ dimensional $\ZZ_2$ descendant of $H_{d+1}$,
we firstly replace the ${d+1}^{th}$ component $p_{d+1}$ of the $d+1$ dimensional momentum with a corresponding lattice form.
More accurately, we replace the $p^2_{d+1}$ in even functions of $p_{d+1}$ with $1-\cos(p_{d+1})$,
$p^2_{d+1}\rightarrow 1-\cos(p_{d+1})$, and replace the $p_{d+1}$ in odd functions of $p_{d+1}$ with $\sin(p_{d+1})$,
$p_{d+1}\rightarrow \sin(p_{d+1})$. We will denote the resulting Hamiltonian
after these replacements as $H_{d+1}(p_{d+1})$.
Then we look the variable $p_{d+1}$ of $H_{d+1}(p_{d+1})$ as an external parameter, rewritten as $\theta$.
Thus, we get a family of $d$ dimensional models $H_{d}(\theta)$ parameterized by $\theta$.
Finally, we set $\theta=0$ and $\theta=\pi$ respectively, to get two specifical $d$ dimensional models,
$H_{d}=H_{d}(0)$ and $H'_{d}=H_{d}(\pi)$. One can easily see that these two $d$ dimensional models are in the same symmetry class
with the parent $H_{d+1}$. The key point is, as demonstrated by \cite{pt1},
if the topological invariant of the parent $H_{d+1}$ is odd,
then one can naturally classify these two $d$ dimensional models $H_{d}$,$H'_{d}$ as
in two different $\ZZ_2$ topological phases. While, if the topological invariant of $H_{d+1}$ is even,
then $H_{d}$ and $H'_{d}$ should be classified as in the same $\ZZ_2$ phase.
Hence, by using this procedure, begin with a $d+1$ dimensional parent model with odd topological invariant,
we always can construct a $d$ dimensional model with nontrivial $\ZZ_2$ topological invariant.

To illustrate this idea, we now give a \emph{theoretical derivation} to the well known model \cite{TI}
for three dimensional topological insulators $\mathrm{Bi_2Se_3}, \mathrm{Bi_2Te_3}$ and $\mathrm{Sb_2Te_3}$.

We begin from a three dimensional model $\{\u(\Bp)=m+t\Bp^2+\p\}$
(here $\Bp=(p_1,p_2,p_3)$, and $\p=p_1\i+p_2\j+p_3\k$) for DIII class.
By using the methods discussed in section 5,
we can construct a dimensional lifted four dimensional class AII model,
\be
H_{\AII}=\left(\begin{array}{ccccc} p_4 & m+t(\Bp^2+p^2_4)+\p
\\ m+t(\Bp^2+p^2_4)-\p & -p_4
\end{array}\right).\label{4d}\ee
We then perform the dimensional reduction procedure with respect to $p_2$
to this 4 dimensional parent model $H_{\AII}$.
Consequently we will get a family of three dimensional models
\be
H_3(\theta)=\left(\begin{array}{ccccc} p_4 & \w(p,\theta)
\\ \bar{\w}(p,\theta) & -p_4
\end{array}\right),\label{3dz2}\ee
here $\w(p,\theta)=m+t(p^2_1+p^2_3+p^2_4)+t(1-\cos(\theta))+p_1\i+p_3\k+\sin(\theta)\j$.

Noticing that, the topological non-triviality of the four dimensional parent model $H_{\AII}$ requires $mt<0$.
Besides, in order to ensure that the procedure of replacing $p_2$
with its corresponding lattice form does not change the topological invariant of $H_{\AII}$,
we need to impose an additional constrain $-2t^2<mt$.

Now, in $H_3(\theta)$ (\ref{3dz2}), we set $\theta=0$ and $\theta=\pi$ respectively,
to get two specifical three dimensional models $H_3(0)$ and $H_3(\pi)$, with $\w(p,0)=m+t(p^2_1+p^2_3+p^2_4)+p_1\i+p_3\k$ and $\w(p,\pi)=m+2t+t(p^2_1+p^2_3+p^2_4)+p_1\i+p_3\k$ respectively.
By construction, these two models must be in different $\ZZ_2$ topological phases.
Choose any one of them as a $\ZZ_2$ trivial model, then the other one must be $\ZZ_2$ nontrivial.
Noticing that $-2t^2<mt<0$,
hence one can see that the model $H_3(\pi)$ can be continuously deformed into an obviously trivial model,
thus one can naturally choose this model as the topologically trivial one.
Then, the other three dimensional model $H_3(0)$ will naturally be $\ZZ_2$ nontrivial.

To rewrite this $\ZZ_2$ nontrivial model more explicitly,
we now take $\i=i\sigma_x, \k=-i\sigma_z$,
and then make a unitary transformation $\left(\begin{array}{ccccc} 1 & 1
\\ 1 & -1\end{array}\right)/\sqrt{2}$, after resetting $p_1=p_z,p_4=p_x, p_3=p_y$,
we will finally get
\be \left(\begin{array}{ccccc} {\mathcal{M}(\Bp)} & 0 & p_+ & -ip_z\\
0 & {\mathcal{M}(\Bp)} & -ip_z & p_- \\
p_- & ip_z & -{\mathcal{M}(\Bp)} & 0\\
ip_z & p_+ & 0 & -{\mathcal{M}(\Bp)}
\end{array}\right),\label{z2ti}\ee
where ${\mathcal{M}(\Bp)}=m+t\Bp^2$, with $\Bp=(p_x,p_y,p_z)$, and $p_\pm=p_x\pm i p_y$.
After reordering the basis appropriately, we can rewrite this three dimensional $\ZZ_2$ nontrivial model as
\be H_{3d}=\left(\begin{array}{ccccc} {\mathcal{M}(\Bp)} & -ip_z & 0 & p_+\\
ip_z & -{\mathcal{M}(\Bp)} & p_+ & 0 \\
0 & p_- & {\mathcal{M}(\Bp)}& -ip_z\\
p_- & 0 & ip_z & -{\mathcal{M}(\Bp)}
\end{array}\right).\label{z2ti}\ee
This is just the well known model for three dimensional topological
insulators $\mathrm{Bi_2Se_3}, \mathrm{Bi_2Te_3}$ and $\mathrm{Sb_2Te_3}$.

One can continue this dimensional reduction one dimension further, to get the second $\ZZ_2$
descendant of the parent $d+1$ dimensional model $H_{d+1}$. For example, we begin from the parent four dimensional model $H_{\AII}$ (\ref{4d}),
and then perform the dimensional reduction with respect to $p_3$.
Then after setting $p_1=p_x,p_2=p_y$, we can get a three dimensional model (with an appropriate unitary transformation)
\be \left(\begin{array}{ccccc} M(p,p_4) & p_4 & 0 & -p_-\\
p_4 & -M(p,p_4) & p_- & 0 \\
0 & p_+ & M(p,p_4) & p_4\\
-p_+ & 0 & p_4 & -M(p,p_4)
\end{array}\right),\ee
where, $M(p,p_4)=m+t(|p|^2+p^2_4)$, with $|p|^2=p^2_x+p^2_y$.
Now, we further reduce this three dimensional model with respect to $p_4$.
After reordering the basis appropriately, we can finally get
\be H_{2d}=\left(\begin{array}{ccccc} M(p) & -p_-  & 0 & 0\\
-p_+ & -M(p) & 0 & 0 \\
0 & 0 & M(p) & p_+\\
0 & 0 & p_- & -M(p)
\end{array}\right),\label{z2ti}\ee
here, $M(p)=m+t|p|^2$. This is just the well
known BHZ model for two dimensional quantum spin Hall insulator.

\section{Class A and Class AIII (The Complex Cases)}

~~~~~Having complete the model constructions for all the topological phases
in the eight real symmetry classes of the periodic table.
We now turn to the model constructions for the remaining two complex symmetry classes,
class A and class AIII. These two classes are relatively simpler than the eight real classes.
The topological insulators of class A are living at even dimensions and have no either $\C$ or $\T$ or $\C\cdot\T$ symmetries.
While the topological insulators of class AIII are living at odd dimensions, and have chiral symmetry only.
Obviously, these two classes of topological insulators can be related by the mechanism discussed in section 5.
Thus we need only to focus on the model constructions for class AIII defined at $d=2k+1$ dimensions.
Then the models for class A at $d=2k+2$ dimensions can be gotten from these class AIII models
through the dimensional lifting mechanism of section 5.

As we will see that the Clifford algebras are also important for the present situation.
But, different from the real cases, the Clifford algebras for these two complex classes are
the Clifford algebras over the field of complex numbers $\CC$ rather than over the field of real numbers $\BR$.
In this situation, the distinguish between $C_d$ and $C'_d$ disappears.
Hence we will denote the corresponding Clifford algebra with $d$ Clifford generators as $C^{c}_d$,
the superscript $c$ stands for the complex field $\CC$.
And we will only consider the complex IRs of $C^{c}_d$.

To construct models for $d=2k+1$ dimensional Class AIII topological insulators, we will also choose the basis
that diagonalize the chiral symmetry, the Hamiltonian will then be off diagonal,\be
\left(\begin{array}{ccccc} 0 & \w(\Bp)
\\ \w^{\dagger}(\Bp) & 0
\end{array}\right).\ee
Moreover, to avoid the appearance of a lot of $i$ in our formulas,
we will choose the Clifford generators $e_i$ of $C^{c}_{2k+1}$ to satisfy $e^2_i=-1$,
although this choice is not essential.
And we will also construct model $\{\w(\Bp)\}$
by using the Clifford element $\w(\Bp)=w_0(\Bp)+w_i(\Bp)e_i$,
but now the components $w_0(\Bp)$ and $w_i(\Bp)$ are in general complex rather than real. In this situation,
the Clifford conjugation $\bar{w}$ is not well defined, but, as long as $w^2_0(\Bp)+\sum_iw^2_i(\Bp)\neq 0$,
we always can define the inverse of $\w(\Bp)$ as $\w^{-1}(\Bp)=(w_0(\Bp)-w_i(\Bp)e_i)/(w^2_0+\sum_iw^2_i)$.
Thus, the topological invariant $\nu_{2k+1}(\w)$ of $\{\w(\Bp)\}$ can also be defined as follows
\be\nu_{2k+1}(\w)&=&\int_{\Bp} \Ch_{2k+1}(\w)\\
&=&\frac{1}{\mathrm{Vol}(S^{2k+1})(2k+1)!}\int_{\Bp}\epsilon^{i_0i_1...i_{2k+1}}\frac{w_{i_0}dw_{i_1}\wedge...\wedge dw_{i_{2k+1}}}
{(w_0^2+\sum_iw^2_i)^{k+1}}\label{se2}\\
&=&\deg(\w).
\ee
Where the integral in the second line (\ref{se2}) is topologically invariance (as one can proof),
and we will still call this integration as the mapping degree of
$\Bp\rightarrow \w(\Bp)$ and denote it as $\deg(\w)$.

To give an explicitly construction for the basic model $\{\u(\Bp)=u_0(\Bp)+u_i(\Bp)e_i\}$, one can simply add a small perturbation with a small even function to
one of the odd functions $u_i(\Bp)$ of our first type of constructions discussed in section 4.2
(or add a small odd function to the even function $u_0(\Bp)$), to break the parities of these functions.
Obviously, this will break the $\C$ symmetry and $\T$ symmetry of the original constructions of section 4.2.
On the other hand, to ensure that the topological invariants are unchanged under these perturbations,
we need to maintain the asymptotic behavior of $\u(\Bp)$ at $\Bp\rightarrow \infty$.
For example, we can take our basic model $\{\u(\Bp)\}$ as
\be (u_0(\Bp),u_1(\Bp),...,u_{d}(\Bp))=(m+t\Bp^2,p_1+\epsilon p^2_1,p_2,...,p_{d}),\ee
here $\epsilon\ll |t|$. Obviously, in this situation, we still have
\be\deg(\u)=\sign(m),\ee
when $mt<0$. And the general models $\{\w(\Bp)\}$ for class AIII can then be constructed as $\{\w(\Bp)=\u^n(\Bp)\}$.

The dimensional lifting of section 5 tells us that the Clifford algebra behind the model constructions
for $2k+2$ dimensional class A topological insulators is $C^{c}_{2k+3}$.
The connection between $2k+1$ dimensional class AIII models and $2k+2$ dimensional class A models relies on
the following isomorphism between the Clifford algebras,
\be C^{c}_{2k+1}\otimes C^{c}_{2}\simeq C^{c}_{2k+3}.\ee
Moreover we would like to note that the mod 2 Bott periodicity of these two complex classes
relies on the mod 2 periodicity of the complex Clifford algebras,
$C^{c}_{2k+1}\otimes \CC(2)\simeq C^{c}_{2k+3}$
(where $\CC(2)\simeq C^{c}_{2}$ stands for the algebra of $2\times 2$ complex matrixes).

As an illustration, we would like to give a \emph{theoretical derivation} for the model for
two dimensional quantum anomalous Hall insulator with the first Chern number $c_1=1$ \cite{qam}\cite{QAr},
since this has been realized in magnetic topological insulator of $\mathrm{Cr}$-doped
$\mathrm{(Bi,Sb)_2Te_3}$\cite{qa}.
For the quantum anomalous Hall insulators with large Chern numbers
we refer the readers to \cite{chern,hc0,hc1,hc2,hc3,hc4}.

We will begin from an one dimensional chiral model (the generator of the Clifford algebra $C^c_1$ is simply the imaginary unit $i$)
\be\left(\begin{array}{ccccc} 0 & m+tp^2_1+p_1i
\\ m+tp^2_1-p_1i & 0
\end{array}\right).\ee
This model can be lifted to a two dimensional model (by using the method discussed in section 5)
\be\left(\begin{array}{ccccc} p_2 & m+t(p^2_1+p^2_2)+p_1i
\\ m+t(p^2_1+p^2_2)-p_1i & -p_2
\end{array}\right).\ee
After an appropriate unitary transformation, this two dimensional model can be transformed to the form
\be H_2=\left(\begin{array}{ccccc} m+t\Bp^2 & p_2+ip_1
\\ p_2-ip_1 & -(m+t\Bp^2)
\end{array}\right),\ee
where $\Bp^2=p^2_1+p^2_2$.
This is just the minimal two bands model \cite{qam}\cite{QAr} for quantum anomalous Hall insulator.\\

{\bf{Acknowledgements}}

The authors acknowledge the support of the Doctoral Startup Package Fund of East
China Institute of Technology (No. DHBK201203).

\end{document}